\documentclass[aps,prl,reprint,superscriptaddress,nofootinbib,floatfix]{revtex4-2}

\usepackage{amsmath,amssymb,bm}
\usepackage{graphicx}
\usepackage{booktabs}
\usepackage{microtype}
\usepackage[hidelinks]{hyperref}

\newcommand{\Tr}{\mathrm{Tr}}
\newcommand{\dd}{\mathrm{d}}
\newcommand{\e}{\mathrm{e}}
\newcommand{\ii}{\mathrm{i}}
\newcommand{\cM}{\mathcal{M}}
\newcommand{\cH}{\mathcal{H}}
\newcommand{\cD}{\mathcal{D}}
\newcommand{\XiC}{\Xi}

\begin{document}

\title{Quantization and Mirror Reduction Do Not Commute in Hamiltonian Embeddings of Nonreciprocal Dynamics}

% Replace author and affiliation information before submission.
\author{Gaurav Sarmah}
\affiliation{Department of Physics and Astronomy, Clemson University, Clemson, SC 29634, USA}
\author{Ramakrishna Podila}
\affiliation{Department of Physics and Astronomy, Clemson University, Clemson, SC 29634, USA}
\email{ramakrp@clemson.edu}

\date{\today}

\begin{abstract}
Hamiltonian embeddings can represent dissipative nonreciprocal classical dynamics exactly on invariant manifolds of enlarged reciprocal systems. We show that canonical quantization of such an embedding need not commute with reduction to the target dynamics. For the mirror construction recently introduced for pairwise nonreciprocal interactions, the invariant manifold is Lagrangian. Exact quantum enforcement of the mirror condition therefore removes the dynamical sector rather than producing a quantum analogue of the reduced flow. If the constraint is imposed only semiclassically, contraction of the target dynamics generates inverse-transpose expansion in the conjugate mirror directions. Quantum uncertainty then gives
\(
\ln\cD\ge \XiC(t)+n\ln[\pi/(\sigma\epsilon)]
\),
where \(\XiC=-\ln|\det M|\) is the accumulated contraction and \(\cD\) the torus Hilbert-space dimension. Exact finite-dimensional Weyl evolution of one- and two-degree-of-freedom nonreciprocal models confirms the resulting logarithmic correspondence time, including its predicted change when the initial localization scales with \(\hbar\). Thus the classical embedding is exact, but its direct canonical quantization is a singular semiclassical construction.
\end{abstract}

\maketitle

\textit{Introduction.---}
Nonreciprocal interactions violate mutual action--reaction at the level of effective degrees of freedom and have emerged as a unifying ingredient across active matter, driven colloids, living systems, robotic and mechanical metamaterials, and open quantum dynamics \cite{Bowick2022,FruchartVitelli2026,Fruchart2021,Ivlev2015,Dinelli2023,Brandenbourger2019,Ghatak2020,Tan2022,McDonald2018,WangClerk2019}. They produce phenomena with no equilibrium counterpart, including nonreciprocal phase transitions, directional defects, traveling and time-dependent ordered states, frustration, and anomalous fluctuation transport \cite{Loos2023,Dadhichi2020,Avni2025,Popli2025,Dopierala2025,Hanai2024,Rouzaire2025,Osat2023}. Their irreversibility and entropy production have consequently become active subjects in nonequilibrium statistical mechanics \cite{ZhangEntropy2023,LoosThermo2020}. A central difficulty is structural: generic nonreciprocal forces do not follow from a single interaction potential and the reduced dynamics need not possess a symplectic Hamiltonian formulation \cite{Ivlev2015,FruchartVitelli2026}.

Shi \emph{et al.} recently introduced a striking resolution for pairwise nonreciprocal classical systems \cite{Shi2026}. By doubling the degrees of freedom they construct a reciprocal Hamiltonian whose dynamics leaves invariant a ``mirror'' manifold; restriction to that manifold exactly reproduces the original nonreciprocal equations, including dissipative XY models with vision-cone and chase-and-run couplings. Unlike an open-system dilation, the auxiliary variables are not traced out. Instead, the target flow is obtained by an invariant classical constraint. This construction makes canonical transformations, Monte Carlo methods, and Floquet Hamiltonian engineering available to systems that originally lacked a Hamiltonian \cite{Shi2026}. It also raises an immediate question identified in Ref.~\cite{Shi2026}: what is the quantum meaning of the embedding itself?

That question sits within a century-old effort to formulate dissipation and nonvariational dynamics in Hamiltonian language. Bateman's doubled oscillator \cite{Bateman1931}, the Caldirola--Kanai construction \cite{Caldirola1941,Kanai1948}, dual-system quantizations \cite{FeshbachTikochinsky1977,ChruscinskiJurkowski2006}, nonconservative variational principles \cite{Galley2013}, contact Hamiltonian mechanics \cite{Bravetti2017}, metriplectic dynamics \cite{MorrisonUpdike2024}, special Poisson structures \cite{Nutku1990}, and Markovian embeddings of memory dynamics \cite{Siegle2010} all restore useful structure by enlarging or modifying the classical description. These approaches also show why quantization is subtle: canonical quantization of dissipative doubled systems can display non-normalizable or resonance-like sectors \cite{ChruscinskiJurkowski2006,Bagarello2019}, while broader non-Lagrangian theories require dedicated quantization frameworks \cite{Kazinski2005,Gitman2007}.

There is, however, no generic obstruction to representing a classical non-Hamiltonian flow in Hilbert space. Koopman's operator formulation \cite{Koopman1931,vonNeumann1932} and its modern Koopman--von Neumann (KvN) extension to nonlinear non-Hamiltonian dynamics \cite{Joseph2020} produce unitary evolution of classical probability amplitudes. Conversely, genuinely quantum non-Hamiltonian dynamics can be constructed directly as completely positive open-system evolution; a recent cascade-quantization scheme gives exact Lindblad generators for broad classes of polynomial nonlinear flows \cite{Chia2025}. Our question is therefore narrower and more physical: \emph{what happens if the canonical auxiliary coordinates of the reciprocal mirror embedding are themselves promoted to quantum observables and the classical mirror condition is retained as the definition of the target sector?}

Here we show that the two operations, canonical quantization and mirror reduction, do not commute. The obstruction is geometric. The mirror manifold of the overdamped embedding is Lagrangian, so treating the mirror equations as strong first-class quantum constraints eliminates the very dynamical coordinates one intended to quantize. Weakening the constraint restores nontrivial states but exposes unavoidable conjugate fluctuations. Symplecticity forces these fluctuations to expand as the inverse transpose of the contracting target tangent flow, yielding a quantitative Hilbert-space bound. We derive the bound for the full Shi construction, verify it exactly in a finite-dimensional Weyl quantization of a directed XY bond, and stress-test its scaling against thresholds, initial states, preparation resources, and nonlinear many-body flows. This is not a no-go theorem for quantizing non-Hamiltonian dynamics; it is a no-go for obtaining the target quantum dynamics by \emph{directly canonically quantizing and then exactly reducing} this class of mirror Hamiltonian embeddings.

\begin{figure}[t]
 \includegraphics[width=\columnwidth]{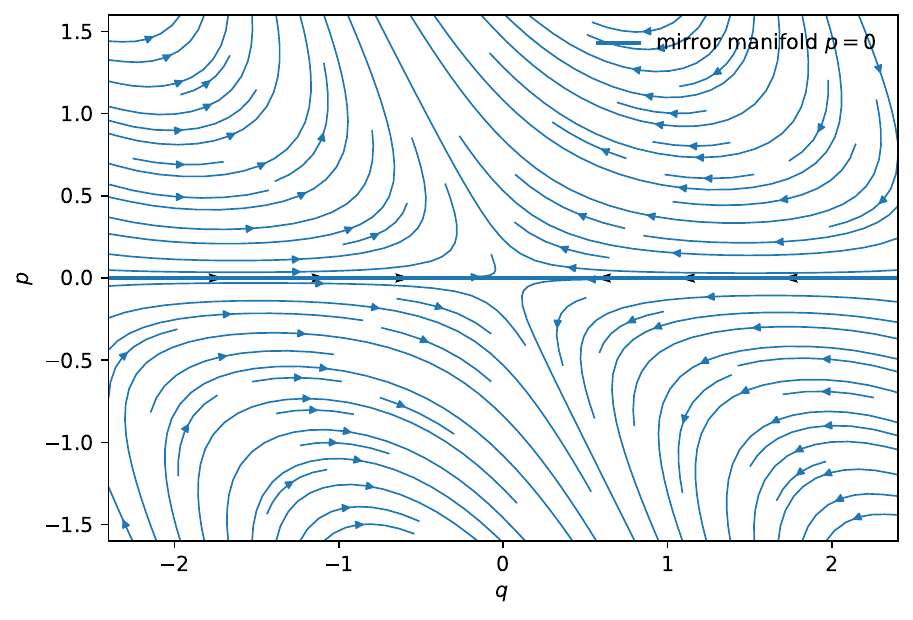}
 \caption{Exact phase flow of the minimal one-way embedded bond, \(H=J[\cos(q+p)-\cos q]\). The line \(p=0\) is the invariant classical mirror manifold. Along it, \(\dot q=-J\sin q\) contracts toward the aligned state, while deviations in the conjugate direction are amplified. The classical reduction is exact only on the Lagrangian manifold itself.}
 \label{fig:phase}
\end{figure}

\textit{Mirror embedding and exact reduction.---}
For the overdamped XY model of Ref.~\cite{Shi2026},
\begin{equation}
 \dot\theta_i=F_i(\bm\theta)
 =-\sum_{j\in N_i}J_{ij}(\theta_i)\sin(\theta_i-\theta_j),
 \label{eq:target}
\end{equation}
with generally \(J_{ij}\neq J_{ji}\). The embedding introduces an auxiliary angle \(\varphi_i\) canonically conjugate to \(\theta_i\), \(\{\theta_i,\varphi_j\}=\delta_{ij}\), and imposes \(\theta_i-\varphi_i=\pi\) \cite{Shi2026}. Define
\begin{equation}
 Q_i=\theta_i,\qquad P_i=\varphi_i-\theta_i+\pi,
 \label{eq:QP}
\end{equation}
so \(\{Q_i,P_j\}=\delta_{ij}\) and the mirror manifold is \(\cM=\{\bm P=0\}\). Combining the system--system and system--auxiliary terms of Ref.~\cite{Shi2026} gives
\begin{equation}
 H=\sum_i\sum_{j\in N_i}J_{ij}(Q_i)
 [\cos(Q_i-Q_j+P_i)-\cos(Q_i-Q_j)].
 \label{eq:exactH}
\end{equation}
Thus \(H(\bm Q,0)=0\), yet \(\partial_{P_i}H|_{\bm P=0}=F_i(\bm Q)\), and Hamilton's equations restricted to \(\cM\) reproduce Eq.~\eqref{eq:target}.

The geometric content is immediate. With \(\omega=\sum_i\dd Q_i\wedge\dd P_i\), \(\dim\cM=n\) in a \(2n\)-dimensional phase space and \(\omega|_{\cM}=0\); hence
\begin{equation}
\cM\ \text{is Lagrangian}.
 \label{eq:lagrangian}
\end{equation}
This matters because a Lagrangian constraint surface is maximally isotropic. If \(P_i=0\) are treated as first-class physical constraints, their characteristic vector fields \(X_{P_i}=\partial_{Q_i}\) span all of \(T\cM\). Standard symplectic reduction therefore quotients away the full tangent dynamics \cite{Dirac1964,AbrahamMarsden1978,MarsdenWeinstein1974,GotayNesterHinds1978}. Equivalently, the Dirac conditions \(\hat P_i|\psi\rangle=0\) imply locally \(\partial_{Q_i}\psi=0\). Refined algebraic quantization and group averaging do not evade this conclusion: averaging over the translations generated by all \(P_i\) projects onto the constant (zero-momentum) sector, which is one-dimensional on the compact torus and one-dimensional after the rigging-map quotient on the noncompact cover \cite{Giulini2000,GiuliniMarolf1999a,GiuliniMarolf1999b}. Exact quantum mirror reduction therefore does not produce a quantum version of Eq.~\eqref{eq:target}; it leaves no local dynamical degree of freedom (Supplemental Material).

The conclusion is nonperturbative on the compact phase space used by Shi \emph{et al.} Each canonical pair is a symplectic two-torus. Quantization of a torus of area \(4\pi^2\) requires \cite{HannayBerry1980,Ligabo2016,Sharatchandra2015}
\begin{equation}
 \hbar_N=\frac{2\pi}{N},\qquad \dim\cH_i=N,
 \label{eq:hbarN}
\end{equation}
with Weyl operators \(U_i=\e^{\ii\hat Q_i}\), \(V_i=\e^{\ii\hat P_i}\) obeying \(U_iV_i=\e^{-\ii\hbar_N}V_iU_i\) in the shift convention specified in the Supplemental Material. Reversing the shift reverses both phase conventions but leaves all geometric and scaling results unchanged. The exact mirror condition \(V_i|\mathrm{mir}\rangle=|\mathrm{mir}\rangle\) selects a one-dimensional eigenspace for each torus. Imposing all mirror conditions thus leaves a one-dimensional joint sector. A quantized Hamiltonian that preserves it acts only by a phase; one that does not preserve it necessarily leaks out of the mirror sector.

\textit{Contraction--uncertainty theorem.---}
For finite-width mirror preparation, expand any smooth embedding with \(H(\bm Q,0)=\mathrm{const}\) and \(\partial_{\bm P}H|_0=\bm F(\bm Q)\) as
\begin{equation}
 H=\bm P^T\bm F(\bm Q)+\frac12\bm P^TB(\bm Q)\bm P+O(P^3).
 \label{eq:genericH}
\end{equation}
For Eq.~\eqref{eq:exactH},
\begin{equation}
 B_{ij}=-\delta_{ij}\sum_kJ_{ik}(Q_i)\cos(Q_i-Q_k).
 \label{eq:B}
\end{equation}
The leading term is the cotangent lift underlying KvN dynamics \cite{Joseph2020}; the \(O(P^2)\) terms distinguish the reciprocal mirror completion and permit transverse fluctuations to feed back into the physical coordinates.

Along a mirror trajectory \(\bm Q_t\), set \(A_t=D\bm F(\bm Q_t)\). Linear fluctuations satisfy
\begin{equation}
 \frac{\dd}{\dd t}
 \begin{pmatrix}\delta\bm Q\\ \delta\bm P\end{pmatrix}
 =
 \begin{pmatrix}A_t&B_t\\0&-A_t^T\end{pmatrix}
 \begin{pmatrix}\delta\bm Q\\ \delta\bm P\end{pmatrix}.
 \label{eq:linear}
\end{equation}
If \(M_t\) is the target tangent propagator, \(\dot M_t=A_tM_t\), then
\begin{equation}
\delta\bm P(t)=M_t^{-T}\delta\bm P(0).
 \label{eq:inverse}
\end{equation}
Thus physical contraction is symplectically compensated by expansion normal to \(\cM\). Defining
\begin{equation}
 \XiC(t)=-\ln|\det M_t|=-\int_0^t\Tr A_s\,\dd s,
 \label{eq:Xi}
\end{equation}
we have
\begin{equation}
 \det\Gamma_P(t)=\e^{2\XiC(t)}\det\Gamma_P(0).
 \label{eq:detP}
\end{equation}
Suppose the initial physical resolution satisfies \(\Gamma_Q(0)\preceq\sigma^2I_n\). The Robertson--Schr\"odinger covariance inequality \cite{Robertson1929} implies \(\det\Gamma\ge(\hbar/2)^{2n}\), while positivity of the block covariance gives \(\det\Gamma\le\det\Gamma_Q\det\Gamma_P\). Therefore
\begin{equation}
 \det\Gamma_P(0)\ge\left(\frac{\hbar}{2\sigma}\right)^{2n}.
 \label{eq:initP}
\end{equation}
Requiring the mirror width to remain \(\Gamma_P(t)\preceq\epsilon^2I_n\) yields
\begin{equation}
\XiC(t)\le n\ln\!\left(\frac{2\sigma\epsilon}{\hbar}\right).
 \label{eq:boundhbar}
\end{equation}
The result does not assume Gaussian states and allows arbitrary initial \(Q\)-\(P\) correlations. Equation~\eqref{eq:inverse} is an exact tangent-flow identity; applying the covariance bound to a finite-width packet requires that it remain in the mirror tube where the quadratic fluctuation expansion is controlled. A sufficient nonlinear exit criterion, with explicit remainder constants, is given in the Supplemental Material. On the torus, \(\cD=N^n\) and Eq.~\eqref{eq:hbarN} give
\begin{equation}
\ln\cD\ge\XiC(t)+n\ln\!\left(\frac{\pi}{\sigma\epsilon}\right).
 \label{eq:Dbound}
\end{equation}
At fixed resolution, every nat of accumulated contraction requires at least one additional nat of Hilbert-space capacity. For \(\XiC\simeq n\kappa t\),
\begin{equation}
 t_*\lesssim\kappa^{-1}\ln N+O(1).
 \label{eq:logtime}
\end{equation}
Logarithmic quantum--classical correspondence times are familiar in chaotic Hamiltonian systems \cite{Silvestrov2002}; here the target flow need not be chaotic. The expanding direction is imposed by the symplectic completion of a contracting reduced flow.

\begin{figure}[t]
 \includegraphics[width=\columnwidth]{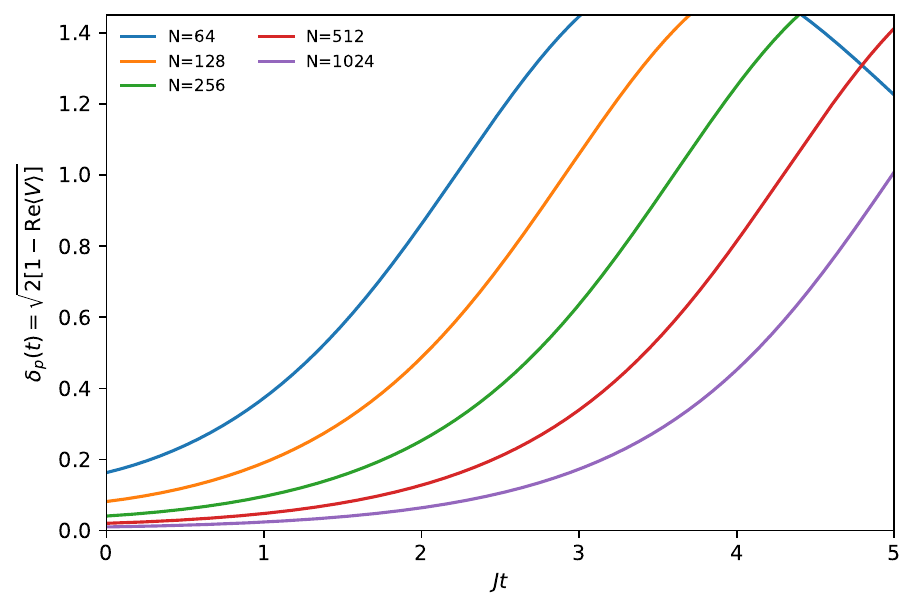}
 \caption{Exact Weyl dynamics for the directed bond with \(J=1\), \(q_0=0.8\), and fixed position width \(\sigma=0.3\). The mirror-width proxy \(\delta_p=\sqrt{2[1-\mathrm{Re}\langle V\rangle]}\) begins as \(O(N^{-1})\) and grows until the finite-\(\hbar_N\) state exits the classical mirror neighborhood. The propagator is evaluated by the direct sparse matrix exponential; no Trotter approximation is used.}
 \label{fig:spread}
\end{figure}

\textit{Exact Weyl test.---}
The minimal directed bond, \(J_{12}=J\), \(J_{21}=0\), closes in relative and mirror coordinates as
\begin{equation}
 H(q,p)=J[\cos(q+p)-\cos q].
 \label{eq:bondH}
\end{equation}
Its exact phase flow is shown in Fig.~\ref{fig:phase}. On \(p=0\), \(\dot q=-J\sin q\) and
\begin{equation}
 \tan\frac{q_{\rm cl}(t)}2=\tan\frac{q_0}2\,\e^{-Jt}.
 \label{eq:qclass}
\end{equation}
Linear transverse deviations obey
\begin{equation}
 \frac{\delta p(t)}{\delta p(0)}
 =\frac{\sin q_0}{\sin q_{\rm cl}(t)}
 =\e^{Jt}\frac{1+a^2\e^{-2Jt}}{1+a^2},\quad a=\tan\frac{q_0}{2}.
 \label{eq:exactgrowth}
\end{equation}
Near alignment, \(H_2=-Jpq-(J/2)p^2\), and exact Heisenberg evolution gives
\begin{equation}
 q(t)=\e^{-Jt}q(0)-\sinh(Jt)p(0).
 \label{eq:qtlin}
\end{equation}
Minimizing \(\mathrm{Var}\,q(t)\) over all allowed initial \(q\)-\(p\) covariances at fixed \(\mathrm{Var}\,q(0)=\sigma^2\) gives
\begin{equation}
\mathrm{Var}\,q(t)\ge\frac{\hbar^2}{4\sigma^2}\sinh^2(Jt),
 \label{eq:exactvar}
\end{equation}
so initial correlations cannot cancel the growing quantum contribution.

\begin{figure}[t]
 \includegraphics[width=\columnwidth]{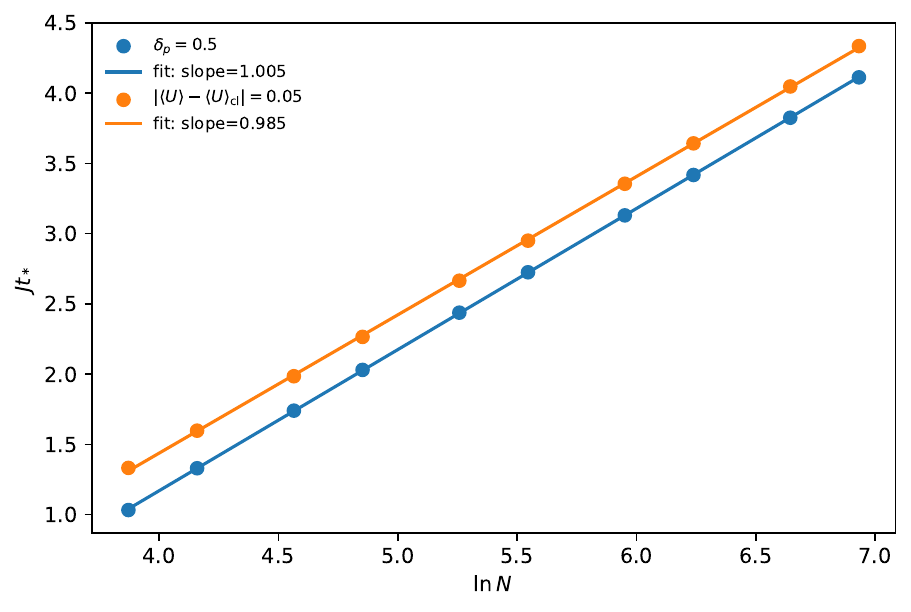}
 \caption{Quantum--classical breakdown times versus \(\ln N\) at fixed initial position resolution. The first mirror-width crossing and the first deviation of the physical Weyl observable \(U=\e^{\ii q}\) from classical Liouville evolution both follow the predicted logarithmic law, with fitted slopes 1.005 and 0.985.}
 \label{fig:scaling}
\end{figure}

We quantize Eq.~\eqref{eq:bondH} with the Weyl pair \(UV=\e^{-\ii\hbar_N}VU\) \cite{HannayBerry1980,Ligabo2016,Keating1991}:
\begin{equation}
 \hat H_N=\frac{J}{2}\left[
 \e^{\ii\hbar_N/2}UV+
 \e^{-\ii\hbar_N/2}V^\dagger U^\dagger-U-U^\dagger\right].
 \label{eq:weylH}
\end{equation}
A periodic Gaussian centered at \(q_0=0.8\) is propagated exactly by \(\exp(-\ii\hat H_Nt/\hbar_N)\). Figure~\ref{fig:spread} shows mirror spreading up to \(N=1024\). To verify that the effect is not confined to the auxiliary coordinate, we compare \(\langle U\rangle_t\) with the classical Liouville average obtained by transporting the same initial probability density under Eq.~\eqref{eq:qclass}. Figure~\ref{fig:scaling} shows the first time at which either \(\delta_p=0.5\) or \(|\langle U\rangle-\langle U\rangle_{\rm cl}|=0.05\). Fits give
\begin{align}
 Jt_*^{(p)}&=(1.0047\pm0.0013)\ln N-2.8489,\nonumber\\
 Jt_*^{(U)}&=(0.9847\pm0.0038)\ln N-2.5009,
 \label{eq:fits}
\end{align}
with \(R^2=0.99999\) and \(0.99989\). The theorem predicts slope unity.

\begin{figure}[t]
 \includegraphics[width=\columnwidth]{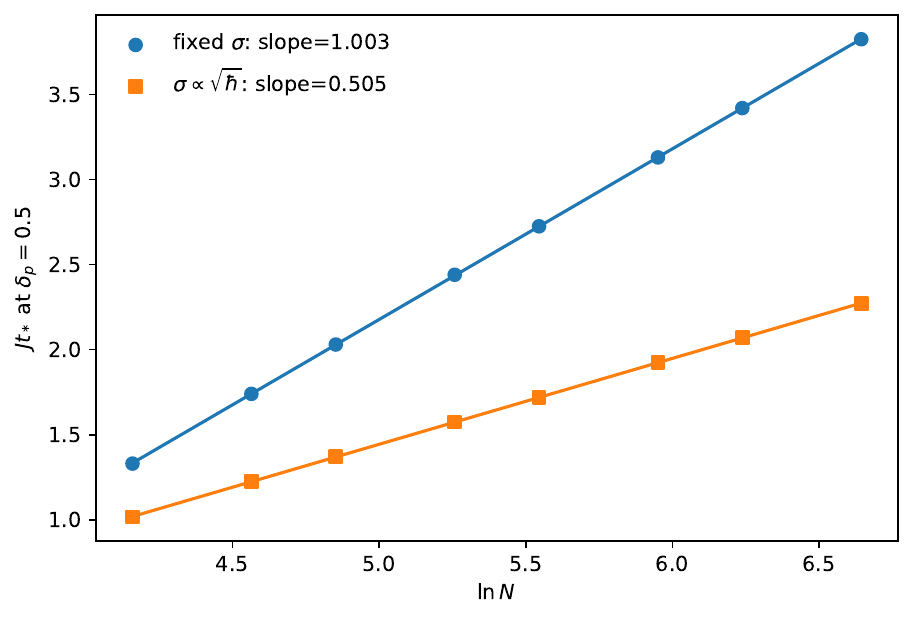}
 \caption{Independent preparation-resource test of Eq.~\eqref{eq:boundhbar}. For fixed \(\sigma\), the breakdown time has slope \(1.0035\) versus \(\ln N\). Choosing \(\sigma_N\propto\sqrt{\hbar_N}\) changes the predicted coefficient from 1 to \(1/2\); exact Weyl dynamics gives 0.5050.}
 \label{fig:prep}
\end{figure}

A stronger falsification test changes the \(\hbar\)-dependence of the initial preparation. At fixed \(\sigma\), minimum mirror uncertainty scales as \(\Delta p_0\sim\hbar_N\), giving the unit slope above. If instead \(\sigma_N\propto\sqrt{\hbar_N}\), then \(\Delta p_0\propto\sqrt{\hbar_N}\), and Eq.~\eqref{eq:boundhbar} predicts that the coefficient of \(\ln N\) is halved. Figure~\ref{fig:prep} shows the same exact Weyl calculation under the two preparations. The measured slopes are 1.0035 and 0.5050, respectively. Threshold scans and initial-condition scans are reported in the Supplemental Material. We additionally perform an exact two-degree-of-freedom quantum Weyl test with asymmetric couplings \(J_{12}=1\), \(J_{21}=0.25\): the contracting relative mode predicts slopes \(1/(J_{12}+J_{21})=0.8\) and \(0.4\) for fixed-width and \(\sigma\propto\sqrt{\hbar}\) preparations, while the exact \(N^2\)-dimensional evolutions give \(0.793\) and \(0.387\), respectively. A nonlinear six-spin directed-ring integration independently verifies Eq.~\eqref{eq:inverse}.

\textit{Discussion.---}
Four distinctions are essential. First, the result does not prohibit quantum descriptions of nonreciprocal or dissipative dynamics. Standard cascaded quantum systems provide an explicit Gorini–Kossakowski–Sudarshan–Lindblad (GKSL) realization of one-way coupling \cite{Gardiner1993,Carmichael1993}, reservoir engineering generalizes this mechanism \cite{MetelmannClerk2015}, and recent cascade and contact-compatible constructions quantize non-Hamiltonian classical structures directly as open dynamics \cite{Chia2025,Colombo2026}. Second, it does not contradict KvN mechanics. The pure cotangent lift \(H=\bm P^T\bm F\) generates the desired classical probability flow even as the auxiliary momentum expands, because \(\dot{\bm Q}=\bm F(\bm Q)\) is independent of \(\bm P\) \cite{Koopman1931,vonNeumann1932,Joseph2020}. In the reciprocal mirror embedding, however, Eq.~\eqref{eq:B} is generically nonzero, so finite transverse quantum width feeds back into the physical coordinates. Third, our geometric obstruction differs from model-specific pathologies of doubled damped oscillators \cite{FeshbachTikochinsky1977,ChruscinskiJurkowski2006,Bagarello2019}. It follows from the geometry of the invariant manifold: the full overdamped construction of Ref.~\cite{Shi2026} is Lagrangian, and its undamped mirror construction is Lagrangian as well (Supplemental Material). Fourth, the conclusion is not asserted for every possible Hamiltonian embedding. If an alternative embedding possessed an invariant symplectic reduced manifold of positive dimension, the dimensional-collapse argument would not apply. The criterion is geometric rather than a blanket statement about nonreciprocity.

The result can be summarized as a noncommuting diagram. Classical reduction of the enlarged reciprocal Hamiltonian gives the target nonreciprocal flow exactly. Canonical quantization of the enlarged system is also well defined. But quantizing first and then imposing the mirror constraint does not yield a nontrivial quantum version of the reduced dynamics: exact reduction collapses the dynamical sector, whereas approximate reduction has an uncertainty-limited lifetime quantified by Eq.~\eqref{eq:Dbound}. For contracting flows, the required canonical Hilbert space grows at least as \(\cD\propto\exp\XiC\). The exact finite-dimensional calculation shows that this is not merely an auxiliary instability; it controls the observable departure from the target Liouville dynamics and changes predictably with the quantum preparation resource.

Hamiltonization can therefore restore classical symplectic structure without supplying a regular route to a physical quantum theory of the reduced dissipative system. In the mirror construction, the classical limit is singular: the invariant manifold on which nonreciprocity is exact has no nontrivial exact quantum counterpart, while finite-width quantum states can shadow it only for a contraction-limited time. This delineates precisely what is gained, and what is not, when a non-Hamiltonian nonreciprocal dynamics is embedded into a reciprocal Hamiltonian phase space.

\clearpage
\onecolumngrid
\appendix
\section*{Supplemental Material for ``Quantization and Mirror Reduction Do Not Commute in Hamiltonian Embeddings of Nonreciprocal Dynamics''}

\section{Exact canonical rewriting of the overdamped mirror Hamiltonian}

The overdamped embedding of Ref.~\cite{Shi2026} contains
\begin{align}
H_{SS}&=-\sum_{\langle ij\rangle}
\left[J_{ij}(\theta_i)+J_{ji}(\theta_j)\right]
\cos(\theta_i-\theta_j),\\
H_{Sa}&=-\sum_{\langle ij\rangle}\left[
J_{ij}(\theta_i)\cos(\theta_j-\varphi_i)
+J_{ji}(\theta_j)\cos(\theta_i-\varphi_j)
\right].
\end{align}
Using \(Q_i=\theta_i\), \(P_i=\varphi_i-\theta_i+\pi\), one has
\begin{equation}
\cos(\theta_j-\varphi_i)
=-\cos(Q_i-Q_j+P_i).
\end{equation}
Collecting terms directed from \(i\) to \(j\) yields Eq.~\eqref{eq:exactH}. Its Taylor series is
\begin{equation}
H=\sum_iP_iF_i(\bm Q)
-\frac12\sum_iC_i(\bm Q)P_i^2
-\frac16\sum_iF_i(\bm Q)P_i^3+O(P^4),
\end{equation}
where
\begin{align}
F_i(\bm Q)&=-\sum_jJ_{ij}(Q_i)\sin(Q_i-Q_j),\\
C_i(\bm Q)&=\sum_jJ_{ij}(Q_i)\cos(Q_i-Q_j).
\end{align}
Thus the leading term is the standard cotangent lift \(\bm P^T\bm F\), while the reciprocal completion generates nonzero curvature in the transverse momenta beginning at \(O(P^2)\).

\section{Lagrangian character and exact reduction}

For \(\omega=\sum_i\dd Q_i\wedge\dd P_i\), the inclusion \(\iota:\cM\hookrightarrow T^{2n}\) with \(\cM=\{P_i=0\}\) gives \(\iota^*\omega=0\). Since \(\dim\cM=n=\frac12\dim T^{2n}\), \(\cM\) is Lagrangian. The constraint functions \(C_i=P_i\) commute, \(\{C_i,C_j\}=0\), and are preserved weakly because
\begin{equation}
\{P_i,H\}=-\partial_{Q_i}H
=-\sum_jA_{ji}P_j+O(P^2)\approx0.
\end{equation}
Their characteristic vector fields are \(X_{P_i}=\partial_{Q_i}\), which span \(T\cM\). Consequently the symplectic quotient of \(\cM\) by the characteristic foliation is a point. This is why treating the mirror condition as an ordinary first-class physical constraint is incompatible with retaining the original \(Q_i\) as dynamical degrees of freedom.

In local Schr\"odinger coordinates, \(\hat P_i=-\ii\hbar\partial_{Q_i}\), and the strong condition \(\hat P_i\psi=0\) forces \(\psi\) to be constant. On the compact torus, the equivalent finite-dimensional statement is that each \(V_i\) has a nondegenerate eigenstate with eigenvalue one, so the joint exact mirror sector is one-dimensional.

\section{Refined algebraic quantization and group averaging}

The conclusion of exact reduction is unchanged if the constraints are imposed by refined algebraic quantization (RAQ) or group averaging rather than by the strong Dirac equation \cite{GiuliniMarolf1999a,GiuliniMarolf1999b,Giulini2000}. For the mirror constraints \(C_i=P_i\), the gauge group generated by \(C_i\) is the group of translations of the \(Q_i\). On a continuum torus, the group-averaging projector is
\begin{equation}
\Pi_0=\frac{1}{(2\pi)^n}\int_{[0,2\pi)^n}\!\dd^n a\,
\exp\!\left(-\frac{\ii}{\hbar}\bm a\cdot\hat{\bm P}\right).
\end{equation}
Acting on \(\psi(\bm Q)\), it returns its spatial average, so \(\mathrm{Ran}\,\Pi_0\) is the one-dimensional constant sector. In the finite Weyl quantization the same statement is discrete:
\begin{equation}
\Pi_0^{(N)}=\prod_{i=1}^n\left(\frac1N\sum_{r=0}^{N-1}V_i^r\right),
\end{equation}
which projects onto the joint \(V_i=1\) eigenspace and has rank one.

On the noncompact cover, group averaging produces distributional zero-momentum solutions rather than normalizable constants. For test states \(\psi,\phi\), the rigging-map inner product is, up to normalization,
\begin{equation}
\eta(\psi)[\phi]=\int_{\mathbb R^n}\dd^na\,
\langle\psi|\e^{-\ii\bm a\cdot\hat{\bm P}/\hbar}|\phi\rangle
\propto \tilde\psi(0)^*\tilde\phi(0).
\end{equation}
After quotienting null vectors and completing, the physical Hilbert space is therefore isomorphic to \(\mathbb C\). Distributional enforcement does not restore \(Q\)-dependent observables because the \(Q\) directions are precisely the gauge orbits generated by the constraints. This is the simple mirror-system realization of the general relation between constrained quantization and reduction \cite{GuilleminSternberg1982}.

\section{Undamped mirror embedding and geometric scope}

The same dimensional obstruction applies to the undamped construction given in the Supplemental Material of Ref.~\cite{Shi2026}. Its ambient canonical variables are \((\theta_i,L_i^\theta;\varphi_i,L_i^\varphi)\), with
\begin{equation}
\omega=\sum_i\left(\dd\theta_i\wedge\dd L_i^\theta+
\dd\varphi_i\wedge\dd L_i^\varphi\right),
\end{equation}
and the invariant constraints are
\begin{equation}
G_i^{(1)}=\theta_i-\varphi_i-\pi=0,\qquad
G_i^{(2)}=L_i^\theta+L_i^\varphi=0.
\end{equation}
They mutually Poisson commute. Parameterizing their common surface by
\begin{equation}
\theta_i=Q_i,\quad\varphi_i=Q_i-\pi,\quad
L_i^\theta=\Pi_i,\quad L_i^\varphi=-\Pi_i,
\end{equation}
gives
\begin{equation}
\iota^*\omega=\sum_i\left(\dd Q_i\wedge\dd\Pi_i+\dd Q_i\wedge\dd(-\Pi_i)\right)=0.
\end{equation}
The constraint surface has dimension \(2n\) in a \(4n\)-dimensional phase space and is therefore Lagrangian. Hence the exact-reduction obstruction is not restricted to the overdamped implementation.

More generally, our exact dimensional-collapse statement applies whenever the target is represented by an invariant Lagrangian mirror manifold whose characteristic distribution spans its tangent bundle. Pairwise interactions are not required by the theorem itself. Conversely, if a different Hamiltonian embedding realizes the target on an invariant \emph{symplectic} manifold, or on a coisotropic manifold whose quotient has positive dimension, exact quantum reduction can retain a nontrivial Hilbert space and our no-go statement need not apply. This geometric criterion specifies the scope without claiming a universal obstruction for all possible embeddings.

\section{Linearized symplectic compensation}

For the general local Hamiltonian
\begin{equation}
H(\bm Q,\bm P)=H_0+\bm P^T\bm F(\bm Q)
+\frac12\bm P^TB(\bm Q)\bm P+O(P^3),
\end{equation}
Hamilton's equations linearized about \(\bm P=0\) give Eq.~\eqref{eq:linear}. Let \(M_t\) solve \(\dot M=A_tM\). Direct differentiation verifies
\begin{equation}
\frac{\dd}{\dd t}M_t^{-T}=-A_t^TM_t^{-T},
\end{equation}
which proves Eq.~\eqref{eq:inverse}. The transverse covariance therefore obeys
\begin{equation}
\Gamma_P(t)=M_t^{-T}\Gamma_P(0)M_t^{-1},
\end{equation}
and hence
\begin{equation}
\det\Gamma_P(t)=\det\Gamma_P(0)|\det M_t|^{-2}.
\end{equation}
The upper-right block \(B_t\) affects \(\delta Q\) through
\begin{equation}
\delta\bm Q(t)=M_t\left[\delta\bm Q(0)+
\int_0^tM_s^{-1}B_sM_s^{-T}\delta\bm P(0)\,\dd s\right].
\label{eq:feedbackS}
\end{equation}
For a pure KvN cotangent lift, \(B=0\); for Eq.~\eqref{eq:exactH}, \(B\) is generically nonzero, so transverse uncertainty contaminates the physical coordinates.

\section{Covariance proof of the quantum resource bound}

Let the full covariance matrix be
\begin{equation}
\Gamma=\begin{pmatrix}\Gamma_Q&C\\C^T&\Gamma_P\end{pmatrix}.
\end{equation}
For \(n\) canonical pairs, the Robertson--Schr\"odinger inequality \(\Gamma+\ii\hbar\Omega/2\succeq0\) implies that every symplectic eigenvalue of \(\Gamma\) is at least \(\hbar/2\), and therefore
\begin{equation}
\det\Gamma\ge(\hbar/2)^{2n}.
\end{equation}
For a positive covariance matrix,
\begin{equation}
\det\Gamma=\det\Gamma_Q\det(\Gamma_P-C^T\Gamma_Q^{-1}C)
\le\det\Gamma_Q\det\Gamma_P.
\end{equation}
If \(\Gamma_Q(0)\preceq\sigma^2I\), then \(\det\Gamma_Q(0)\le\sigma^{2n}\), giving Eq.~\eqref{eq:initP}. If \(\Gamma_P(t)\preceq\epsilon^2I\), then \(\det\Gamma_P(t)\le\epsilon^{2n}\). Combining with Eq.~\eqref{eq:detP} proves Eq.~\eqref{eq:boundhbar}.

The proof allows arbitrary initial cross-covariance \(C\), mixed states, and non-Gaussian states. Its dynamical input is the linearized fluctuation evolution. Hence it is exact for quadratic embeddings and asymptotically controlled for localized semiclassical states until the fluctuation expansion ceases to be valid.

\section{Explicit nonlinear validity window}

The inverse-transpose relation in Eq.~\eqref{eq:inverse} is an exact statement about the tangent flow. To control a finite-width packet, write deviations from a constrained reference trajectory as \(z=(\delta\bm Q,\bm P)\) and the full classical Hamiltonian vector field as
\begin{equation}
\dot z=L(t)z+R(z,t),\qquad \|R(z,t)\|\le K_2\|z\|^2
\label{eq:remainderbound}
\end{equation}
throughout a tube \(\|z\|\le r\). Let \(\Phi_L(t,s)\) be the propagator of the linear problem and assume the conservative bound
\begin{equation}
\|\Phi_L(t,s)\|\le G\e^{\lambda(t-s)},\qquad t\ge s.
\end{equation}
For \(z_0=\|z(0)\|\), define
\begin{equation}
\eta(t)=\frac{G^2K_2z_0}{\lambda}\left(\e^{\lambda t}-1\right),
\label{eq:eta}
\end{equation}
with the continuous \(\lambda\to0\) limit \(\eta=G^2K_2z_0t\). Bihari's inequality applied to the variation-of-constants formula gives the sufficient bound
\begin{equation}
\|z(t)\|\le\frac{Gz_0\e^{\lambda t}}{1-\eta(t)}
\label{eq:nonlinearbound}
\end{equation}
whenever \(\eta(t)<1\) and the right-hand side remains below \(r\). If \(z_L(t)=\Phi_L(t,0)z(0)\), Duhamel's formula additionally gives
\begin{equation}
\|z(t)-z_L(t)\|
\le Gz_0\e^{\lambda t}\frac{\eta(t)}{[1-\eta(t)]^2}.
\label{eq:linearerror}
\end{equation}
Thus a simple conservative error-controlled window is
\begin{equation}
\eta(t)\ll1,\qquad
\frac{Gz_0\e^{\lambda t}}{1-\eta(t)}<r.
\label{eq:exitcriterion}
\end{equation}
Here \(K_2\) can be chosen from a supremum of the second derivatives of the Hamiltonian vector field in the mirror tube, and therefore explicitly grows when the transverse curvature block \(B(Q)\) or higher derivatives become large. For localized semiclassical states, \(z_0\) may be taken as a chosen high-probability localization radius. Covariance alone does not compactly support an arbitrary non-Gaussian state, so Eqs.~\eqref{eq:remainderbound}--\eqref{eq:exitcriterion} are a sufficient finite-width criterion rather than an assertion about arbitrarily heavy tails. The exact Weyl calculations in the Letter do not use this approximation and directly test the predicted asymptotic scaling beyond the conservative local estimate.

\section{Unidirectional bond: analytic growth and optimized uncertainty}

For Eq.~\eqref{eq:bondH}, Hamilton's equations are
\begin{align}
\dot q&=-J\sin(q+p),\\
\dot p&=J[\sin(q+p)-\sin q].
\end{align}
On \(p=0\), Eq.~\eqref{eq:qclass} follows. Linearizing the second equation in \(p\),
\begin{equation}
\dot{\delta p}=J\cos[q_{\rm cl}(t)]\delta p.
\end{equation}
Using \(\dd\ln(\sin q_{\rm cl})/\dd t=-J\cos q_{\rm cl}\) gives Eq.~\eqref{eq:exactgrowth}.

Near \(q=p=0\), the quadratic Hamiltonian \(H_2=-Jpq-(J/2)p^2\) is understood with symmetric ordering. Its Heisenberg equations are identical to the classical linear system, producing Eq.~\eqref{eq:qtlin}. Write \(a=\e^{-Jt}\), \(b=-\sinh(Jt)\), \(\mathrm{Var}(q_0)=\sigma^2\), \(\mathrm{Cov}(q_0,p_0)=c\). The uncertainty relation gives
\begin{equation}
\mathrm{Var}(p_0)\ge\frac{c^2+\hbar^2/4}{\sigma^2}.
\end{equation}
Therefore
\begin{equation}
\mathrm{Var}(q_t)\ge a^2\sigma^2+
\frac{b^2}{\sigma^2}(c^2+\hbar^2/4)+2abc.
\end{equation}
Minimizing over \(c\) gives \(c=-a\sigma^2/b\) and Eq.~\eqref{eq:exactvar}. Thus even optimally correlated initial states cannot eliminate the growing term.

For fixed \(q\)-resolution \(\sigma\), an initially minimum-uncertainty mirror width scales as \(\Delta p_0\sim\hbar_N/(2\sigma)=\pi/(N\sigma)\). From Eq.~\eqref{eq:exactgrowth}, a fixed threshold \(\epsilon\) is crossed asymptotically at
\begin{equation}
Jt_*=\ln N+\ln\frac{\epsilon\sigma}{\pi}
-2\ln\cos\frac{q_0}{2}+o(1).
\label{eq:intercept}
\end{equation}
If \(\sigma\propto\sqrt{\hbar_N}\), then \(\Delta p_0\propto\sqrt{\hbar_N}\) and the slope is \(1/2\).

\section{Exact finite-dimensional Weyl dynamics}

We choose the \(N\)-point position grid \(q_m=-\pi+2\pi m/N\), with
\begin{align}
U|q_m\rangle&=\e^{\ii q_m}|q_m\rangle,\\
V|q_m\rangle&=|q_{m-1}\rangle,
\end{align}
so \(UV=\e^{-\ii\hbar_N}VU\). With this convention, for integers \(a,b\), symmetric Weyl ordering is
\begin{equation}
\operatorname{Op}_{W}[\e^{\ii(aq+bp)}]
=\e^{\ii ab\hbar_N/2}U^aV^b.
\label{eq:weylconvention}
\end{equation}
In particular, \(\operatorname{Op}_{W}[\e^{\ii(q+p)}]=\e^{\ii\hbar_N/2}UV\), which gives Eq.~\eqref{eq:weylH}. Choosing instead \(V|q_m\rangle=|q_{m+1}\rangle\) changes \(UV=\e^{+\ii\hbar_N}VU\) and simultaneously reverses the phase in Eq.~\eqref{eq:weylconvention}; it is only a convention. Other admissible orderings with the same principal symbol alter subprincipal \(O(\hbar_N)\) leakage coefficients but cannot change the rank-one exact mirror sector, the classical tangent identity, or the asymptotic contraction--uncertainty scaling. The initial wavefunction is
\begin{equation}
\psi_m(0)=\mathcal N
\exp\left[-\frac{d_{2\pi}(q_m-q_0)^2}{4\sigma^2}\right],
\end{equation}
where \(d_{2\pi}\) is the shortest periodic distance. All dynamics reported in the Letter use the sparse Krylov action of the exact matrix exponential \(\exp(-\ii\hat H_Nt/\hbar_N)\); the time increment \(\Delta t=0.0025/J\) is only the sampling interval and introduces no Trotter error.

The mirror-width observable is
\begin{equation}
\delta_p(t)=\sqrt{2[1-\mathrm{Re}\langle V\rangle_t]},
\end{equation}
which equals the rms momentum angle to leading order for a narrow distribution around \(p=0\). For the physical benchmark we use \(U=\e^{\ii q}\). The classical Liouville value is evaluated without time stepping:
\begin{equation}
\langle U\rangle_{\rm cl}(t)=
\sum_m|\psi_m(0)|^2\e^{\ii q_{\rm cl}(t;q_m)},
\end{equation}
where \(q_{\rm cl}\) is given by Eq.~\eqref{eq:qclass}.

The numerical values underlying Figs.~\ref{fig:spread} and \ref{fig:scaling} are
\begin{center}
\begin{tabular}{rrrrr}
\toprule
\(N\)&\(\hbar_N\)&\(\delta_p(0)\)&\(t_p\)&\(t_U\)\\
\midrule
48&0.130900&0.216875&1.0325&1.3325\\
64&0.098175&0.163079&1.3300&1.5975\\
96&0.065450&0.108921&1.7400&1.9850\\
128&0.049087&0.081744&2.0300&2.2650\\
192&0.032725&0.054521&2.4375&2.6650\\
256&0.024544&0.040898&2.7250&2.9500\\
384&0.016362&0.027268&3.1300&3.3550\\
512&0.012272&0.020452&3.4175&3.6425\\
768&0.008181&0.013635&3.8250&4.0475\\
1024&0.006136&0.010226&4.1125&4.3350\\
\bottomrule
\end{tabular}
\end{center}
Here \(t_p\) is the first crossing of \(\delta_p=0.5\), and \(t_U\) is the first crossing of \(|\langle U\rangle-\langle U\rangle_{\rm cl}|=0.05\), in units of \(J^{-1}\).

\section{Stress tests}

\subsection{Threshold robustness}
For \(N=64,96,128,192,256,384,512,768\), \(q_0=0.8\), and \(\sigma=0.3\), fitting \(Jt_*=s\ln N+b\) gives
\begin{center}
\begin{tabular}{rrrr}
\toprule
\(\delta_p\) threshold & \(s\) & \(b\) & \(R^2\)\\
\midrule
0.3&1.011855&-3.435994&0.999942\\
0.4&1.005384&-3.094791&0.999978\\
0.5&1.003498&-2.840233&0.999993\\
0.6&1.000403&-2.614763&0.999997\\
0.8&1.000263&-2.259631&0.999997\\
\bottomrule
\end{tabular}
\end{center}

\subsection{Initial-condition robustness}
At fixed threshold \(\delta_p=0.5\) and \(\sigma=0.3\),
\begin{center}
\begin{tabular}{rrrrr}
\toprule
\(q_0\)&s&fitted \(b\)&asymptotic \(b\) from Eq.~\eqref{eq:intercept}&\(R^2\)\\
\midrule
0.3&1.000403&-2.949763&-3.019265&0.999997\\
0.8&1.003498&-2.840233&-2.877392&0.999993\\
1.2&1.005244&-2.654659&-2.657920&0.999983\\
1.6&1.006874&-2.370966&-2.319068&0.999981\\
\bottomrule
\end{tabular}
\end{center}
The slope is insensitive to the initial point, while the intercept follows the analytic dependence on \(q_0\).

\subsection{Coherent-packet scaling}
For \(\sigma_N=\sqrt{\hbar_N/2}\), the predicted slope is \(1/2\). The simulations give
\begin{center}
\begin{tabular}{rrrr}
\toprule
threshold&slope&intercept&\(R^2\)\\
\midrule
0.5&0.504985&-1.081343&0.999971\\
0.8&0.497023&-0.478965&0.999981\\
\bottomrule
\end{tabular}
\end{center}
This confirms that the logarithmic coefficient is set by the \(\hbar\)-scaling of the initial localization resource, as implied by Eq.~\eqref{eq:boundhbar}.

\section{Fully quantum two-degree-of-freedom Weyl test}

To test the scaling beyond a single canonical pair, we quantize the coupled asymmetric two-spin mirror Hamiltonian
\begin{align}
H={}&J_{12}[\cos(Q_1-Q_2+P_1)-\cos(Q_1-Q_2)]\nonumber\\
&+J_{21}[\cos(Q_2-Q_1+P_2)-\cos(Q_2-Q_1)],
\label{eq:twoDOFH}
\end{align}
with \(J_{12}=1\) and \(J_{21}=0.25\). On the mirror manifold,
\begin{equation}
\dot Q_1=-J_{12}\sin(Q_1-Q_2),\qquad
\dot Q_2=+J_{21}\sin(Q_1-Q_2).
\end{equation}
The relative coordinate \(q=Q_1-Q_2\) obeys
\begin{equation}
\dot q=-(J_{12}+J_{21})\sin q,
\end{equation}
so the asymptotic contracting exponent is \(\kappa=J_{12}+J_{21}=1.25\). The theorem therefore predicts
\begin{equation}
t_*^{\rm fixed}=\kappa^{-1}\ln N+O(1)=0.8\ln N+O(1),
\end{equation}
and, for \(\sigma_N\propto\sqrt{\hbar_N}\),
\begin{equation}
t_*^{\sqrt\hbar}=(2\kappa)^{-1}\ln N+O(1)=0.4\ln N+O(1).
\end{equation}

\begin{figure}[t]
\centering
\includegraphics[width=0.68\textwidth]{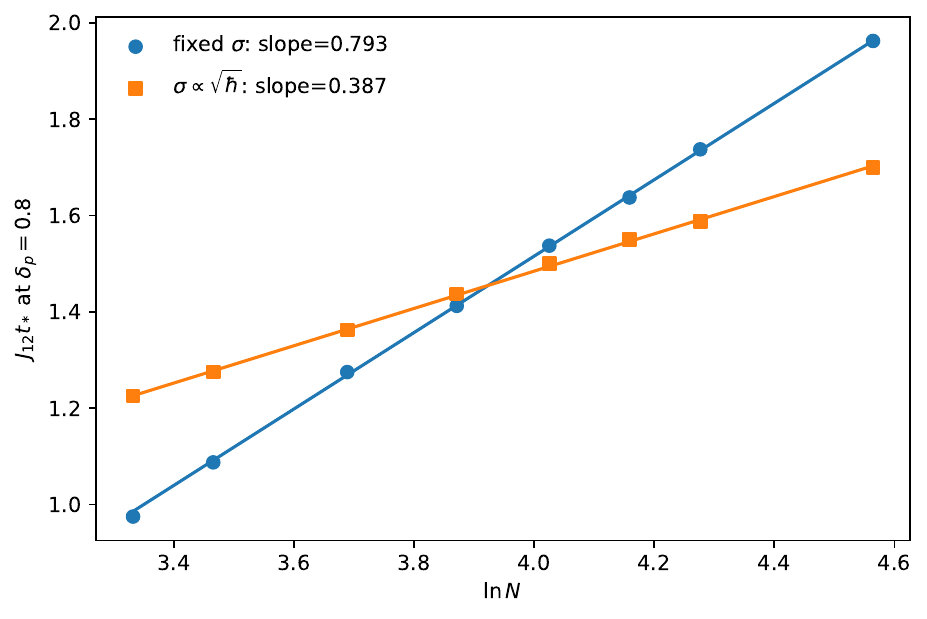}
\caption{Exact two-degree-of-freedom Weyl test of Eq.~\eqref{eq:twoDOFH}. The mirror-breakdown time is linear in \(\ln N\). Fixed position resolution gives slope 0.793, close to the asymptotic prediction \(1/(J_{12}+J_{21})=0.8\); choosing \(\sigma_N\propto\sqrt{\hbar_N}\) gives slope 0.387, close to the predicted 0.4. No Trotter approximation is used.}
\label{fig:twoDOF}
\end{figure}

We use the tensor-product Weyl Hilbert space of dimension \(N^2\), the convention in Eq.~\eqref{eq:weylconvention}, product packets centered at \((Q_1,Q_2)=(0.6,-0.4)\), and exact sparse-matrix exponential propagation. Define
\begin{equation}
\delta_{p,\mathrm{rms}}=
\left\{\frac12\sum_{i=1}^2 2[1-\mathrm{Re}\langle V_i\rangle]\right\}^{1/2}.
\end{equation}
Using the first crossing of \(\delta_{p,\mathrm{rms}}=0.8\), exact calculations for \(N=28\)--96 give a fixed-width slope \(0.7934\) (fit for \(N\ge32\), \(R^2=0.99984\)) and a \(\sigma_N=\sqrt{\hbar_N/2}\) slope \(0.3873\) (\(R^2=0.99949\)), in agreement with the predicted \(0.8\) and \(0.4\). Figure~\ref{fig:twoDOF} shows the result. Thus the resource-dependent logarithmic law survives in a genuinely coupled \(N^2\)-dimensional quantum problem.

\begin{center}
\begin{tabular}{rrrrr}
\toprule
\(N\)&\(N^2\)&\(\hbar_N\)&\(t_*^{\rm fixed}\)&\(t_*^{\sqrt\hbar}\)\\
\midrule
28&784&0.22440&0.9750&1.2250\\
32&1024&0.19635&1.0875&1.2750\\
40&1600&0.15708&1.2750&1.3625\\
48&2304&0.13090&1.4125&1.4375\\
56&3136&0.11220&1.5375&1.5000\\
64&4096&0.09817&1.6375&1.5500\\
72&5184&0.08727&1.7375&1.5875\\
96&9216&0.06545&1.9625&1.7000\\
\bottomrule
\end{tabular}
\end{center}

\section{Nonlinear many-body check of the inverse-transpose identity}

To verify Eq.~\eqref{eq:inverse} away from the analytically soluble bond, we integrate a six-site directed XY ring with
\begin{equation}
F_i=-J_R\sin(Q_i-Q_{i+1})-J_L\sin(Q_i-Q_{i-1}),
\end{equation}
with \(J_R=1\), \(J_L=0.2\), and initial condition
\begin{equation}
\bm Q_0=(0.2,-0.4,0.7,-0.1,0.5,-0.6).
\end{equation}
Simultaneously integrating
\begin{equation}
\dot M=A M,\qquad \dot N=-A^TN,
\end{equation}
with \(M(0)=N(0)=I\) to \(t=5\) gives
\begin{equation}
\max_t\frac{\|N(t)-M(t)^{-T}\|_F}{\|M(t)^{-T}\|_F}
=5.6\times10^{-11},
\end{equation}
while the accumulated contraction reaches \(\XiC(5)=35.3298\). This numerical check is not needed for the proof, but verifies the implementation for a nonlinear many-body directed flow.

\section{Exact mirror state under Weyl quantization}

Let \(|m\rangle=N^{-1/2}\sum_k|q_k\rangle\), so that \(V|m\rangle=|m\rangle\). For Eq.~\eqref{eq:weylH},
\begin{equation}
\langle m|\hat H_N|m\rangle=0,
\qquad
\|\hat H_N|m\rangle\|=
\sqrt{2}J\left|\sin\frac{\hbar_N}{4}\right|.
\end{equation}
Hence its survival probability begins as
\begin{equation}
|\langle m|\e^{-\ii\hat H_Nt/\hbar_N}|m\rangle|^2
=1-\frac{2J^2\sin^2(\hbar_N/4)}{\hbar_N^2}t^2+O(t^4),
\end{equation}
whose coefficient tends to \(J^2/8\) as \(N\to\infty\). Thus the most natural Weyl quantization does not preserve the exact one-dimensional mirror sector. Operator-ordering counterterms could be chosen to force invariance of that sector, but any Hamiltonian restricted to a one-dimensional sector produces only a phase and cannot reproduce the target \(q\) dynamics. The dimensional obstruction is therefore ordering independent even though the short-time leakage coefficient is not.

\section{Scope, KvN comparison, and a positive GKSL counterpart}

The bound does not apply to a KvN encoding in which \(P\) is an auxiliary generator and no physical state is required to have simultaneous resolution in \(Q\) and \(P\). For the pure cotangent lift
\begin{equation}
H_{\rm KvN}=\bm P^T\bm F(\bm Q),
\end{equation}
the transverse equation is still \(\dot{\delta P}=-A^T\delta P\), but the off-diagonal block is exactly \(B=0\). Hence the potentially expanding auxiliary generator does not feed back into \(\dot Q=F(Q)\). By contrast, the reciprocal mirror Hamiltonian in Eq.~\eqref{eq:exactH} has the explicit nonzero block in Eq.~\eqref{eq:B}; this is why finite physical mirror width matters. The distinction is between a Hilbert-space encoding of a classical probability flow and a canonical quantization in which the mirror coordinates themselves are treated as physical conjugate observables.

Nor does the obstruction prohibit open quantum nonreciprocity. A concrete positive counterpart is the standard cascaded GKSL construction \cite{Gardiner1993,Carmichael1993}. For two modes or qubits with lowering operators \(c_1,c_2\), define
\begin{align}
L&=\sqrt{\gamma_1}c_1+\sqrt{\gamma_2}c_2,\\
H_{\rm cas}&=\frac{\ii}{2}\sqrt{\gamma_1\gamma_2}
(c_1^\dagger c_2-c_2^\dagger c_1),
\end{align}
and
\begin{equation}
\dot\rho=-\ii[H_{\rm cas},\rho]+\mathcal D[L]\rho,
\qquad
\mathcal D[L]\rho=L\rho L^\dagger-\tfrac12\{L^\dagger L,\rho\}.
\label{eq:cascadeGKSL}
\end{equation}
For linear modes the adjoint equations are
\begin{equation}
\dot{\langle c_1\rangle}=-\frac{\gamma_1}{2}\langle c_1\rangle,
\qquad
\dot{\langle c_2\rangle}=-\frac{\gamma_2}{2}\langle c_2\rangle
-\sqrt{\gamma_1\gamma_2}\langle c_1\rangle,
\end{equation}
so system 1 drives system 2 without reciprocal back-action. Taking \(c_i=\sigma_i^-\) gives the standard cascaded two-qubit realization; reservoir-engineered coherent/dissipative interference generalizes the same principle \cite{MetelmannClerk2015}. Cascade quantization \cite{Chia2025} and recent contact-compatible Lindblad constructions \cite{Colombo2026} likewise illustrate that open-system quantization can be regular even when direct mirror reduction is not. Equation~\eqref{eq:cascadeGKSL} is not claimed to be the unique quantum analogue of Eq.~\eqref{eq:target}; it is included to make explicit what our no-go statement does \emph{not} exclude.

Piecewise-defined vision-cone couplings have switching surfaces at which \(D\bm F\) is discontinuous. Equations \eqref{eq:inverse}--\eqref{eq:Dbound} apply on smooth trajectory segments and concatenate across nonsingular switching events by multiplying the corresponding tangent maps. The exact one-way bond, the two-degree-of-freedom quantum test, and the smooth directed-ring test avoid this technicality.

\section{Data and reproducibility}
The numerical data, exact-Weyl simulation script, and stress-test tables used in the manuscript accompany this draft. The matrix exponential is evaluated directly with sparse Krylov methods; no hardware noise, fitting of dynamical parameters, or stochastic sampling is involved.

\end{document}